\documentclass[12pt]{iopart}
\usepackage{iopams}
\usepackage{epsf}

\begin{document}

\noindent 
{\scriptsize Published in Journal of Physics G: Nuclear and Particle
Physics. Copyright 1999 IOP Publishing Ltd}

\title[Numerical analysis of triplet pair production cross-sections]{On the numerical analysis of triplet pair production cross-sections and the
mean energy of produced particles for modelling electron-photon cascade in a
soft photon field}

\author{V Anguelov\dag , S Petrov\dag , L Gurdev\ddag\ and J Kourtev\ddag}

\address{\dag\ Institute for Nuclear Research and Nuclear Energy, 
Bulgarian Academy of Sciences, 72 Tzarigradsko Chaussee, 1784 Sofia, Bulgaria}

\address{\ddag\ Institute of Electronics,
Bulgarian Academy of Sciences, 72 Tzarigradsko Chaussee, 1784 Sofia, Bulgaria}

\begin{abstract}
The double and single differential cross-sections with respect to positron
and electron energies as well as the total cross-section of triplet
production in the laboratory frame are calculated numerically in order to
develop a Monte Carlo code for modelling electron-photon cascades in a soft
photon field. To avoid numerical integration irregularities of the
integrands, which are inherent to problems of this type, we have used
suitable substitutions in combination with a modern powerful program code (%
\textit{Mathematica}) allowing one to achieve reliable higher-precission
results. The results obtained for the total cross-section closely agree with
others estimated analytically or by a different numerical approach. The
results for the double and single differential cross-sections turn out to be
somewhat different from some reported recently. The mean energy of the
produced particles, as a function of the characteristic collisional
parameter (the electron rest frame photon energy), is calculated and
approximated by an analytical expression that revises other known
approximations over a wide range of values of the argument. The
primary-electron energy loss rate due to triplet pair production is shown to
prevail over the inverse Compton scattering loss rate at several ($\sim$2)
orders of magnitude higher interaction energy than that predicted formerly.
\end{abstract}

\section{Introduction}

There are two main reasons why triplet pair production (TPP) has been
commonly ignored in astrophysical applications. The first reason is that TPP
is a third-order QED process. The second reason is the extremely complicated
and long expression for the total differential cross-section \cite{mork71}.
This considerably complicates the modelling of the energies and momenta of
the final three particles in comparison with the case of considering only
the two major processes for electron-photon cascade in a radiation field:
pair production and inverse Compton scattering. Apart from the formal
complications there are serious mathematical problems to be overcome
connected with numerical calculation of the double differential
cross-section (DDCS) and single differential cross-section (SDCS) in the
laboratory frame. A typical problem is the integration over the cosine ($%
\cos \theta _{-}$) of the polar angle $\theta _{-}$ of the produced electron
momentum $p_{-}$, where both the integrand irregularities coincide with the
integration limits whose semi-vicinities provide the major contribution to
the integral.

Despite the above-mentioned arguments, at ultrarelativistic electron
energies TPP becomes a prevailing process playing an important part in the
electron-photon cascades in a soft background photon field that form the
energy spectrum from a variety of astrophysical sources. This fact has been
recently emphasized and confirmed by incorporating TPP into full cascade
calculations \cite{mast86}-\cite{mast94}.

Based on the recently revived interest in a more precise simulation of
electron-photon cascades in a photon field, we have started to develop a
Monte Carlo code for modelling TPP. Our intention is to use this code for a
more detailed study of the development of electromagnetic cascades in
thermal fields. To realize this intention we have decided to follow a method
like that suggested in paper \cite{mast91}. Thus, as an initial step we have
precisely recalculated the DDCS and SDCS with respect to electron and
positron energies as well as the total cross-section of TPP in the
laboratory frame. An obstacle to overcome here is the presence of the
above-mentioned integrand irregularities in combination with extremely short
integration intervals. So, one purpose of the present study is to search for
ways to avoid these intrinsic difficulties and to achieve more precise
results. Another purpose of the study is to calculate and analytically
approximate the mean energy of particles produced as a function of the
characteristic collisional parameter. Investigating the primary-electron TPP
energy loss rate is also an important aim of the work.

\section{Method and results}

\subsection{Calculation approach}

Let us first consider the basic expressions of interest corrected \cite
{angu95} for typographical errors that have appeared in many papers. The
DDCS with respect to the positron and electron energies is given by

\begin{eqnarray}
\fl\frac{\rmd^2\sigma }{\rmd E_{+}\rmd E_{-}}(E_0,\varepsilon
_0,s,E_{+},E_{-})  \nonumber \\
\lo =\frac{\alpha _f\,r_0^2}{4\pi ^2s}p_{+}p_{-}\int_{x_{\min }}^1\rmd %
x\int_{y_{\min }(x)}^{y_{\max }(x)}\rmd y[a_1(y_{\max }-y)(y-y_{\min
})]^{-1/2}\int_0^{2\pi }X\rmd \phi _{+},  \label{eq1}
\end{eqnarray}
where $p_{+}$, $E_{+}$, $\theta _{+}$ and $p_{-}$, $E_{-}$, $\theta _{-}$
are momenta, energies and polar angles of the produced positron and
electron, respectively, $x=\cos \theta _{+}$ and $y=\cos \theta _{-}$, $\phi
_{+}$ is the azimuthal angle of the positron, $E_0$ and $\varepsilon _0$ are
the energies of the incoming electron and photon respectively, $%
s=E_0\varepsilon _0(1-\beta \cos \theta )$ is the characteristic collisional
parameter ($\theta $ is the collision angle) representing the photon energy
in the electron rest frame (ERF), and $X$ is a cumbersome expression that is
given in the appendix; the quantities $\alpha _f=e^2/(\hbar c)$ and $%
r_0=e^2/(m_ec^2)$ are the fine structure constant and the classical electron
radius, respectively, and $m_e$ is the electron rest mass. The limits of
integration are:

\begin{figure}[tbp]
\begin{center}
\mbox{\epsfysize=2.6in \epsffile{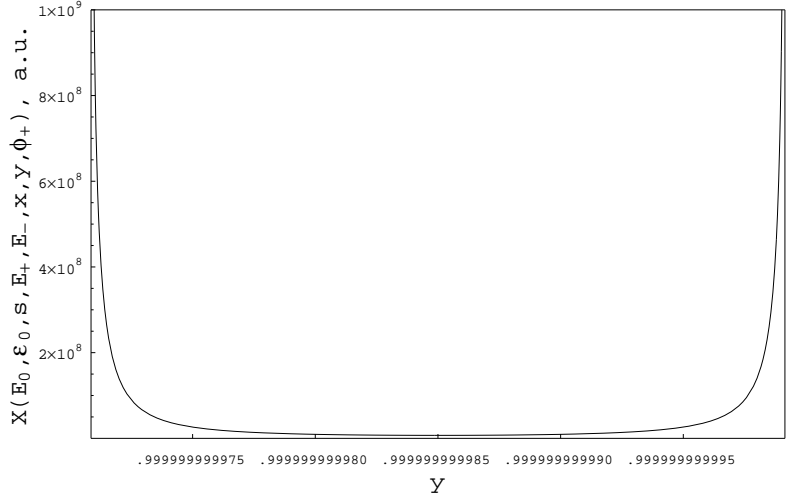}}
\end{center}
\caption{Integrand function $X(E_0,\varepsilon _0,s,E_{+},E_{-},x,y,\phi
_{+})[a_1(y_{\max }-y)(y-y_{\min })]^{-1/2}$ in arbitrary units (before
change of variables) versus the polar-angle cosine $y=\cos \theta _{-}$ at $%
E_0=10^8$, $\varepsilon _0=10^{-3}$, $s=10^5$, $E_{+}=4.999925000425\times
10^7$, $E_{-}=2.500037499837\times 10^7$, $x=\cos \theta _{+}=0.999999999995$%
, $\phi _{+}=\pi $.}
\label{fig:Integrand1}
\end{figure}

\[
y_{\max }=[b_1+(b_1^2-a_1c_1)^{1/2}]/a_1, 
\]

\[
y_{\min }=[b_1-(b_1^2-a_1c_1)^{1/2}]/a_1, 
\]

\[
x_{\min }=-(F_1+p_{-}F_2^{1/2})/(p_{+}P_{tot}), 
\]
where 
\[
\fl a_1=p_{-}^2(P_{tot}^2+p_{+}^2-2P_{tot}p_{+}x),\qquad
b_1=Ap_{-}(p_{+}x-P_{tot}), 
\]

\[
\fl c_1=A^2-p_{+}^2p_{-}^2\sin ^2\theta _{+}, 
\]

\[
\fl A=1+s+p_{+}P_{tot}x-E_{tot}(E_{+}+E_{-})+E_{+}E_{-}, 
\]

\[
\fl F_1=E_{-}^2-(E_{tot}-E_{+})E_{-}+s-E_{tot}E_{+}, 
\]

\[
\fl F_2=(E_{+}+E_{-}-E_{tot})^2-1,\qquad E_{tot}=E_0+\varepsilon _0,\qquad 
\mathrm{and}\qquad \overrightarrow{P}_{tot}=\overrightarrow{p}_0+%
\overrightarrow{k}. 
\]
$\overrightarrow{p}_0$ and $\overrightarrow{k}$are the momenta of the
incoming electron and photon, respectively. Let us note that throughout this
paper the energy quantities are in units $m_ec^2$. Note that the integrand
in equation (\ref{eq1}) has two irregular points with respect to the
variable $y$. These points, $y=y_{\min }$ and $y=y_{\max }$, coincide with
the integration limits. Also, the integration intervals over $x$ and $y$,
especially at high energies, are extremely short. In addition, even within
such narrow integration intervals the integrand changes sharply with
changing of $y$ (see figure 1). Because of these peculiarities of the
integrand, an accurate calculation (numerical integration) can be successful
only when a sufficiently high-precision number is used. This was first been
pointed out by Mastichiadis \cite{mast91} who underlined the necessity of
quadratic precision to calculate DDCS and SDCS, and reconsidered his own
formerly obtained results \cite{mast86}. Since our purpose here is to
perform similar calculations with a higher precision (e.g. up to 80
significant decimal digits) we have used the program code \textit{%
Mathematica }\cite{wolf96}, which allows one to work with arbitrary
high-precision numbers. Thus, one can both eliminate the precision number
conditioning and ensure a reliable precision of the final results. In
addition, this code has an adaptive program for quadrature of multiple
integrals that precisely approximates the fast-changing integrand and
permits one to obtain results with a prescribed precision. Nevertheless, the
extraordinary character of the integrand in equation (\ref{eq1}) leads to a
fast growth of the CPU time with the increase of the interaction energies.
Besides, it is not unknown for the program to fail in some cases. As a
result of searching for ways to eliminate the above-mentioned problems, we
came to the following change of variables that led to acceptable integration
intervals and acceptable smooth behaviour of the integrand:

\begin{figure}[tbp]
\begin{center}
\mbox{\epsfysize=2.6in \epsffile{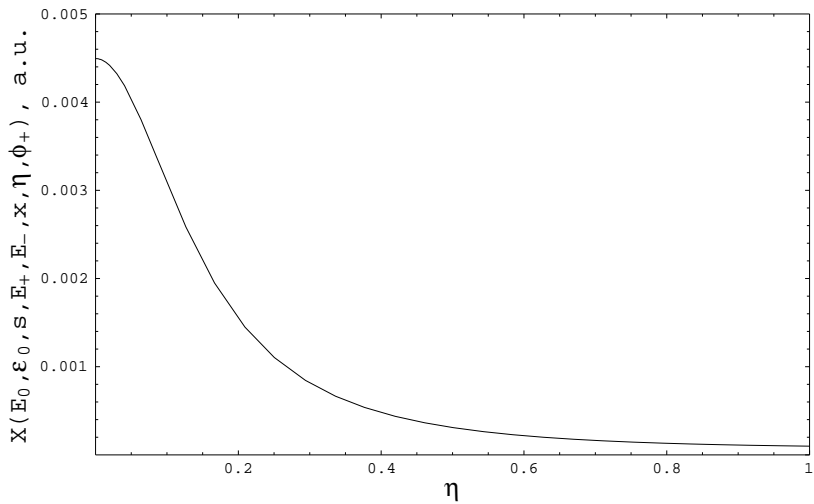}}
\end{center}
\caption{Integrand function $2X(E_0,${$\varepsilon _0,s,$}$%
E_{+},E_{-},x,\eta ,\phi _{+})/(1+\eta ^2)$ in arbitrary units (after change
of variable $y$) versus the new variable $\eta $ at the same fixed values of
the remaining variables as in figure 1.}
\label{fig:Integrand2}
\end{figure}

\numparts
\begin{eqnarray}
x &=&x(\xi )=\xi /l+x_{\min }\quad (l>1),  \label{eq2a} \\
y &=&y(\xi ,\eta )=[y_{\max }(x)+y_{\min }(x)\eta ^2]/(1+\eta ^2){.}
\label{eq2b}
\end{eqnarray}
\endnumparts 

Equation (\ref{eq2b}) is, in fact, one of the Euler substitutions that is
appropriate for this case and leads to the integral:

\begin{eqnarray}
\fl I=I(E_0,\varepsilon _0,s,E_{+},E_{-})=2l^{-1}\int_0^{l(1-x_{\min })}\rmd%
\xi \int_0^\infty \rmd\eta  \nonumber \\
\lo\times \int_0^{2\pi }\rmd\phi _{+}X(E_0,\varepsilon _0,s,E_{+},E_{-},\xi
,\eta ,\phi _{+})/(1+\eta ^2){;}  \label{eq3}
\end{eqnarray}

\[
\fl \rmd^2\sigma /(\rmd E_{+}\rmd E_{-})=(\alpha _f\,r_0^2/(4\pi
^2a_1^{1/2}s))p_{+}p_{-}I(E_0,\varepsilon _0,s,E_{+},E_{-}){.} 
\]

\begin{figure}[tbp]
\begin{center}
\mbox{\epsfysize=3.0in \epsffile{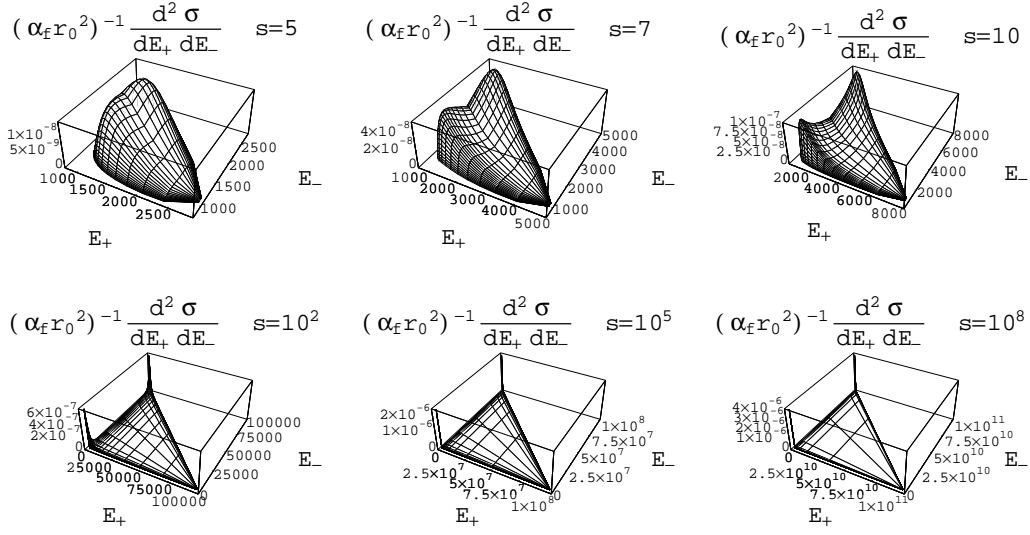}}
\end{center}
\caption{DDCS $(\alpha _fr_0^2)^{-1}\rmd^2\sigma /(\rmd E_{+}\rmd E_{-})$ as
a function of $E_{+}$ and $E_{-}$ at various values of $s $; $\varepsilon
_0=10^{-3}$, $E_0=s/\varepsilon _0$ (glancing collision).}
\label{fig:DDCS1}
\end{figure}

\begin{figure}[tbp]
\begin{center}
\mbox{\epsfysize=2.2in \epsffile{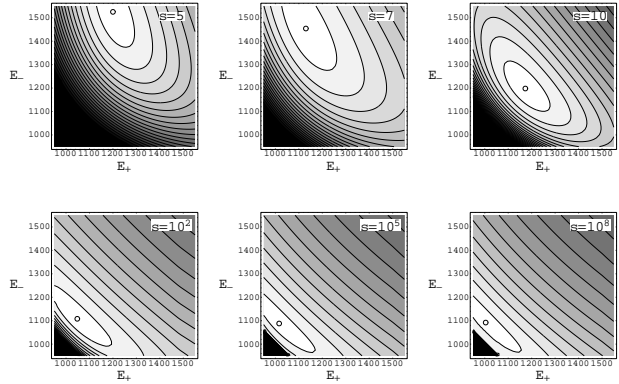}}
\end{center}
\caption{Three-dimensional contour plot of $(\alpha _fr_0^2)^{-1}\rmd%
^2\sigma /(\rmd E_{+}\rmd E_{-})$ as a function of $E_{+}$ and $E_{-}$, at
values of $s$ as in figure 3, respectively, around the first peak of figure
3 disposed near $E_{+,\min }$ and $E_{-,\min }$; $\varepsilon _0=10^{-3}$, $%
E_0=s/\varepsilon _0$ (glancing collision).}
\label{fig:DDCS2}
\end{figure}

It can be seen that the new variables lead to an enlargement of the
integration scale and removal of the integrand irregularities. As one may
expect, the integrand becomes a smooth function of $\eta $ (figure 2), which
leads to an acceleration of the calculation procedure, increase of the
precision, and elimination of any computational failures when \textit{%
Mathematica} is used.

\begin{figure}[tbp]
\begin{center}
\mbox{\epsfysize=2.4in \epsffile{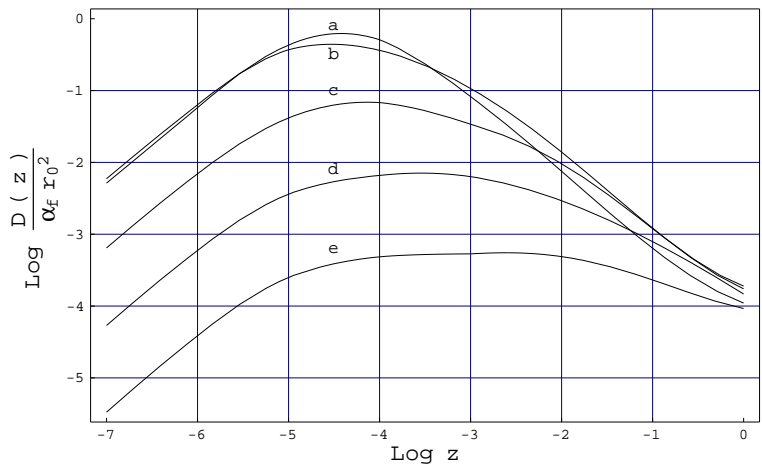}}
\end{center}
\caption{Dependence of the quantity $(\alpha
_fr_0^2)^{-1}D(s,z)=(E_{-}-E_{-,\min })(E_{+}-E_{+,\min })[\rmd^2\sigma /( 
\rmd E_{+}\rmd E_{-})]$ on the parameter $z=(E_{-}-E_{-,\min
})/(E_{-,med}-E_{-,\min })$ [$E_{-,med}=0.5(E_{-,\min }+E_{-,\max })$] for $%
E_{-,\min }<E_{-}<E_{-,med}$ and various outgoing positron energies $%
E_{+}=10^3$($a$),$10^4$($b$),$10^5$($c$),$10^6$($d$) and $10^7$($e$); $%
E_0=10^8$ and $\varepsilon _0=10^{-3}$ ($s=10^5$).}
\label{fig:DDPC}
\end{figure}

\subsection{ Double differential cross-section}

We have obtained precise results for DDCS as a function of $E_0$, $%
\varepsilon _0$, $s$, $E_{+}$ and $E_{-}$. Some of these results are shown
in figures 3 and 4 in a three-dimensional surface form and a
three-dimensional contour form, respectively. It is seen (figure 3) that the
dependence of $\rmd^2\sigma /(\rmd E_{+}\rmd E_{-})$ on $E_{+}$ and $E_{-}$
considered over the whole range of the arguments $E_{+}$ and $E_{-}$, at
various fixed values of the parameter $s=\varepsilon _0E_0$ ( in the case of
a glancing collision when $\theta =\pi /2$ ), is represented by a
double-peak surface whose secants with the planes $E_{+}=const$ are
symmetric curves with respect to the point $(E_{+},E_{-,med})$, where $%
E_{-,med}=0.5(E_{-,\min }+E_{-,\max })$. With the increase of $s$, the
surface peak heights increase, and the surface itself (as well as the
corresponding dependence of DDCS on $E_{+}$ and $E_{-}$) becomes sharper.
Also, up to values of $s=10^2$, the peaks change their positions over the
plane $\{E_{+},E_{-}\}$. So, the common coordinate of both peaks along $%
E_{+} $ axis is shifted to lower values of $E_{+}$. In addition, the first
peak (disposed below the point $(E_{+},E_{-,med})$) is shifted to lower
values of $E_{-}$ (see figure 4), and the second one, to higher values of $%
E_{-}$. At values of $s$ increasing above $s=10^2$ the peaks do not change
their positions; the surface is as if consisting of two spikes (with a
common $E_{+}$ coordinate) whose positions are close to the points $%
(E_{+,\min },E_{-,\min })$ and $(E_{+,\min },E_{-,\max })$, respectively.
The same results as above are represented in figure 5, but in a parametrized
form proposed in \cite{mast91}, where the quantity $D(s,z)=(E_{-}-E_{-,\min
})(E_{+}-E_{+,\min })[\rmd^2\sigma /(\rmd E_{+}\rmd E_{-})]$ is considered
as a function of the parameter $z=(E_{-}-E_{-,\min })/(E_{-,med}-E_{-,\min
}) $ at fixed values of $E_{+}$ and $s$. It is shown in \cite{mast91} that $%
D $ can be considered as dependent only on $s$ and $z$ if $E_0\gg 1\gg
\varepsilon _0$. Such a parametrization is useful for application to a Monte
Carlo code for modelling electron-photon cascades in a soft photon field
taking into account the contribution of the TPP process. Then, on the basis
of tabulated data $D(s,z)$ one can determine DDCS at any combination of $E_0$%
, $\varepsilon _0$, $s$ and positron energy $E_{+}$ \cite{mast91}.
Comparison between our results and those obtained in \cite{mast91} shows
that our curves pass through maximum and that at lower electron energies
tending to $E_{-,\min }$, where the calculation is sensitive to loss of
precision, they essentially fall below the corresponding curves obtained in 
\cite{mast91}.

\begin{figure}[tbp]
\begin{center}
\mbox{\epsfysize=2.4in \epsffile{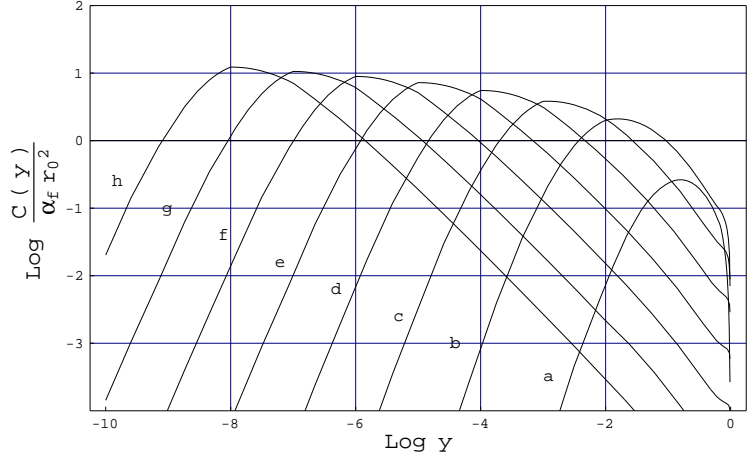}}
\end{center}
\caption{Dependence of the quantity $(\alpha
_fr_0^2)^{-1}C(s,y)=(E_{+}-E_{+,\min })[\rmd\sigma /\rmd E_{+}]$ on the
parameter $y=(E_{+}-E_{+,\min })/(E_{+,\max }-E_{+,\min })$ for various ERF
energies of collision $s=10$($a$),$10^2$($b$),$10^3$($c$),$10^4$($d$),$10^5$(%
$e$),$10^6$($f$),$10^7$($g$) and $10^8$($h$).}
\label{fig:SDPC}
\end{figure}

\begin{table}[tbp]
\caption{TPP total cross-sections (in units $\alpha _f\,r_0^2$).}
\label{tab1}%
\begin{indented}
\lineup
\item[]\begin{tabular}{@{}*{5}{l}}
\br                              
$s,s_{\perp}$&$\sigma _{tot,Haug}$&$\sigma _{tot,Mast}$&$\sigma _{tot,Our}$&$\sigma _{toti,Our}$\cr 
\mr
4.01&$5.80\times 10^{-7}$&$5.4\times 10^{-7}$&$5.8\times 10^{-7}$&0.092\cr
4.1 &$7.53\times 10^{-5}$&$7.6\times 10^{-5}$&$7.5\times 10^{-5}$&0.102\cr 
5   &0.0170&0.019&0.0170&0.221\cr 
7   &0.179 &0.19&0.179 &0.56 \cr
10  &0.594 &0.59&0.59  &1.12 \cr
$10^{2}$&7.21&7.3&7.3 &7.9  \cr
$10^{3}$&15.1&15.3&15.1&15.8 \cr
$10^{4}$&22.6&22.4&22.7&23.2 \cr
$10^{5}$&29.9&29.7&29.9&30.5 \cr
$10^{6}$&37.1&37.0&37. &38.  \cr
$10^{7}$&44.2&-&44. &45.  \cr
$10^{8}$&51.4&-&52. &52.  \cr 
\br
\end{tabular}
\end{indented}
\end{table}

\subsection{Single differential cross-section}

The SDCS $\rmd\sigma /\rmd E_{+}$ has been calculated by integrating $\rmd%
^2\sigma /(\rmd E_{+}\rmd E_{-})$ over $E_{-}$ with integration limits $%
E_{\_}|_{\min }^{\max }=0.5[E_{tot}-E_{+}\pm (P_{tot}-p_{+})(1-2/B)^{1/2}]$, 
$(B=1+s-E_{tot}E_{+}+P_{tot}p_{+})$. In the same way, $\rmd\sigma /\rmd %
E_{-} $ has been calculated by integrating $\rmd^2\sigma /(\rmd E_{+}\rmd %
E_{-})$ over $E_{+}$. The use of an optimum-power spline technique allowed
us to obtain precise results for sufficiently high values of $s>10^8$. (We
consider as an optimum power of the spline that one, above which the results
from the integration remain stable.) The same results have been obtained by
direct integration (without spline interpolation) but using considerably
longer CPU time. Some of the results obtained are shown in figure 6 in a
parametrized form where the quantity $C(s,y)=(E_{+}-E_{+,\min })[\rmd\sigma /%
\rmd E_{+}]$ is considered as a function of the parameter $%
y=(E_{+}-E_{+,\min })/(E_{+,\max }-E_{+,\min })$ at fixed values of $s$. For 
$E_0\gg 1\gg \varepsilon _0$, $C(s,y)$ depends only on $s$ and $y$, and
(when tabulated) allows one to determine the SDCS for any combination of $%
E_0 $, $\varepsilon _0$ and $s$ \cite{mast91}. The comparison between our
results and those obtained in \cite{mast91} shows that the corresponding
curves have the same behaviour with a characteristic maximum. The difference
is that the left-hand part of each curve of ours (including the maximum),
where the calculation is more sensitive to loss of precision, is as if
shifted right with respect to the analogous part of the corresponding curve
obtained in \cite{mast91}.

\subsection{Total cross-section}

We have also calculated the total cross-section $\sigma _{tot}$ by
integrating $\rmd\sigma /\rmd E_{+}$ over $E_{+}$. The integration limits
are: $E_{+}|_{\min }^{\max }=\{E_{tot}(s-1)\pm
P_{tot}[s(s-4)]^{1/2}\}/(1+2s) $ \cite{mast91}. The results obtained are
compared, in table \ref{tab1}, with the results obtained by other authors 
\cite{mast91}, \cite{haug81}. The agreement is excellent and may be
considered as an indirect confirmation of the precise character of our
calculations.

\subsection{Mean energy of produced particles}

Finally, we have calculated the mean energy $E_m=E_{+,m}=E_{-,m}$ of the
produced particles on the basis of the relations:

\begin{eqnarray}
E_m &=&E_{+,-,m}=\int E_{+,-}(\rmd\sigma /\rmd E_{+,-})\rmd E_{+,-}/\sigma
_{tot}  \label{eq4} \\
&&  \nonumber
\end{eqnarray}
The results obtained for $E_{+,m}$ and $E_{-,m}$ are practically coincident,
which may be considered as another confirmation of the precise character of
the calculations performed. On the basis of the parametrization approach
developed in \cite{mast91} we can show that the ratio $E_m/E_0$ is a
function only of $s$ when $E_0\gg 1\gg \varepsilon _0$. In this case, the
integration limits $E_{+,\min }$ and $E_{+,\max }$ (see above) are
expressible as $E_{+,\min }=E_0f_{\min }(s)$ and $E_{+,\max }=E_0f_{\max
}(s) $, where the functions $f_{\min }$ and $f_{\max }$ depend only on $s$.
After the change of variable $E_{+}=xE_0$, instead of equation (\ref{eq4})
we can write:

\begin{eqnarray}
E_m &=&E_0\sigma _{tot}^{-1}(s)\int_{f_{\min }(s)}^{f_{\max }(s)}\rmd x.x.[(%
\rmd\sigma /\rmd x)(E_0,\varepsilon _0,s,xE_0)]{.}  \label{eq5} \\
&&  \nonumber
\end{eqnarray}

In the same way, taking into account that $\sigma _{tot}$ depends only on $s$%
, we obtain

\begin{eqnarray}
\fl \sigma _{tot}(s) =\int_{f_{\min }(s)}^{f_{\max }(s)}\rmd x[(\rmd\sigma /%
\rmd x)(E_0,\varepsilon _0,s,xE_0)]=\int_{f_{\min }(s)}^{f_{\max }(s)}\rmd %
x[(\rmd\sigma /\rmd x)(E_0^{^{\prime }},\varepsilon _0^{\prime
},s,xE_0^{\prime })]  \label{eq6} \\
&&  \nonumber
\end{eqnarray}
where $E_0^{^{\prime }}$ and $\varepsilon _0^{\prime }$ are some other
values of the incoming electron and photon energies, respectively, but such
that the value of $s$ is retained. Based on equation (\ref{eq6}) we may
conclude that $(\rmd\sigma /\rmd x)(E_0^{^{\prime }},\varepsilon _0^{\prime
},s,xE_0^{\prime })=(\rmd\sigma /\rmd x)(E_0,\varepsilon _0,s,xE_0)$, and
consequently (see equation (\ref{eq5})) $E_m/E_0=E_m^{^{\prime
}}/E_0^{^{\prime }}=f(s)$, where

\begin{eqnarray}
f(s) &=&\sigma _{tot}^{-1}(s)N(s){,}  \label{eq7} \\
&&  \nonumber
\end{eqnarray}
and

\begin{eqnarray}
N(s)=\int_{f_{\min }(s)}^{f_{\max }(s)}\rmd x.x.[(\rmd\sigma /\rmd %
x)(E_0,\varepsilon _0,s,xE_0)]{.}  \label{eq8} \\
&&  \nonumber
\end{eqnarray}

\begin{figure}[tbp]
\begin{center}
\begin{tabular}{c}
\epsfysize=2.2in\epsffile{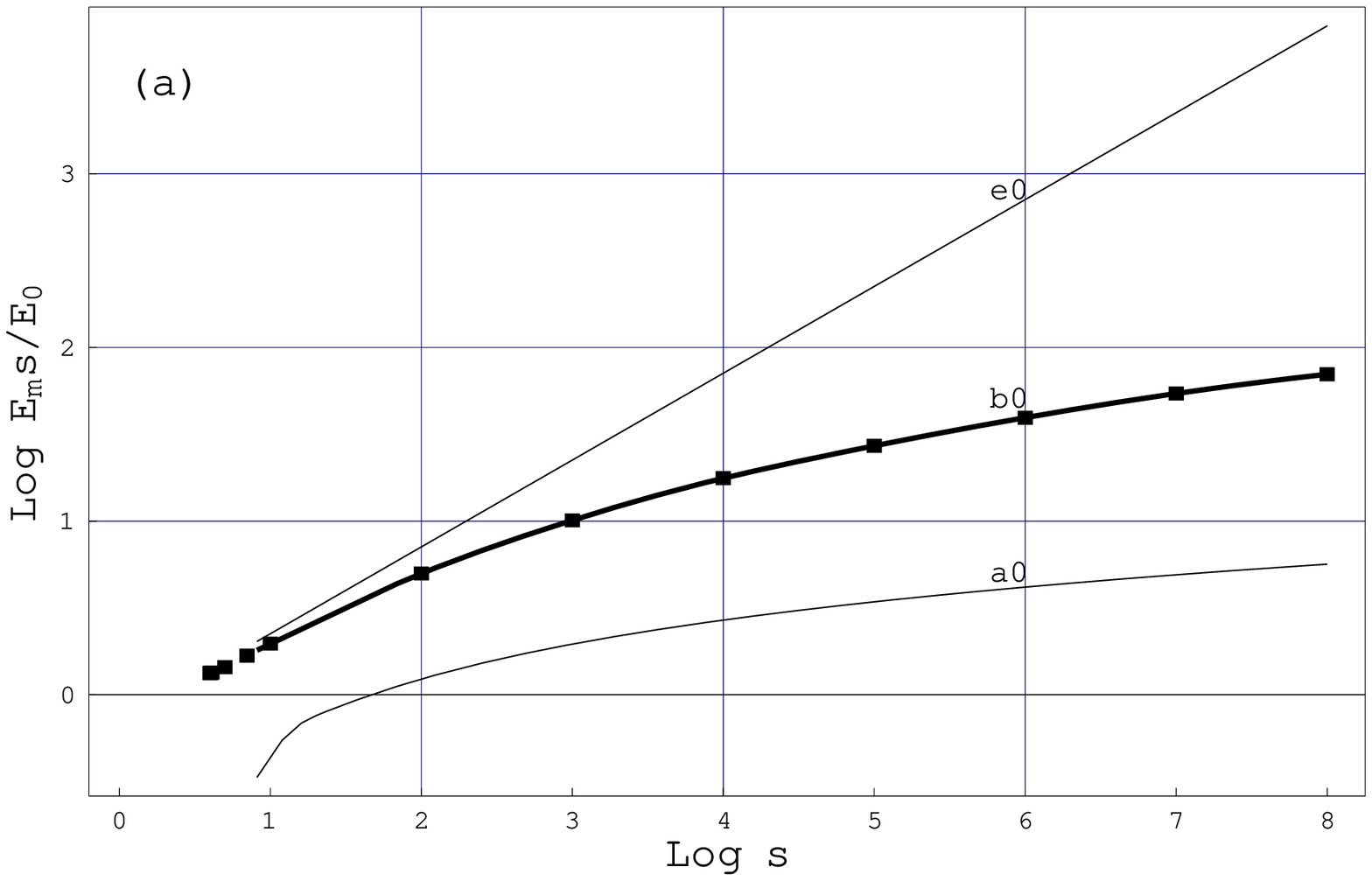} \\ 
\epsfysize=2.2in\epsffile{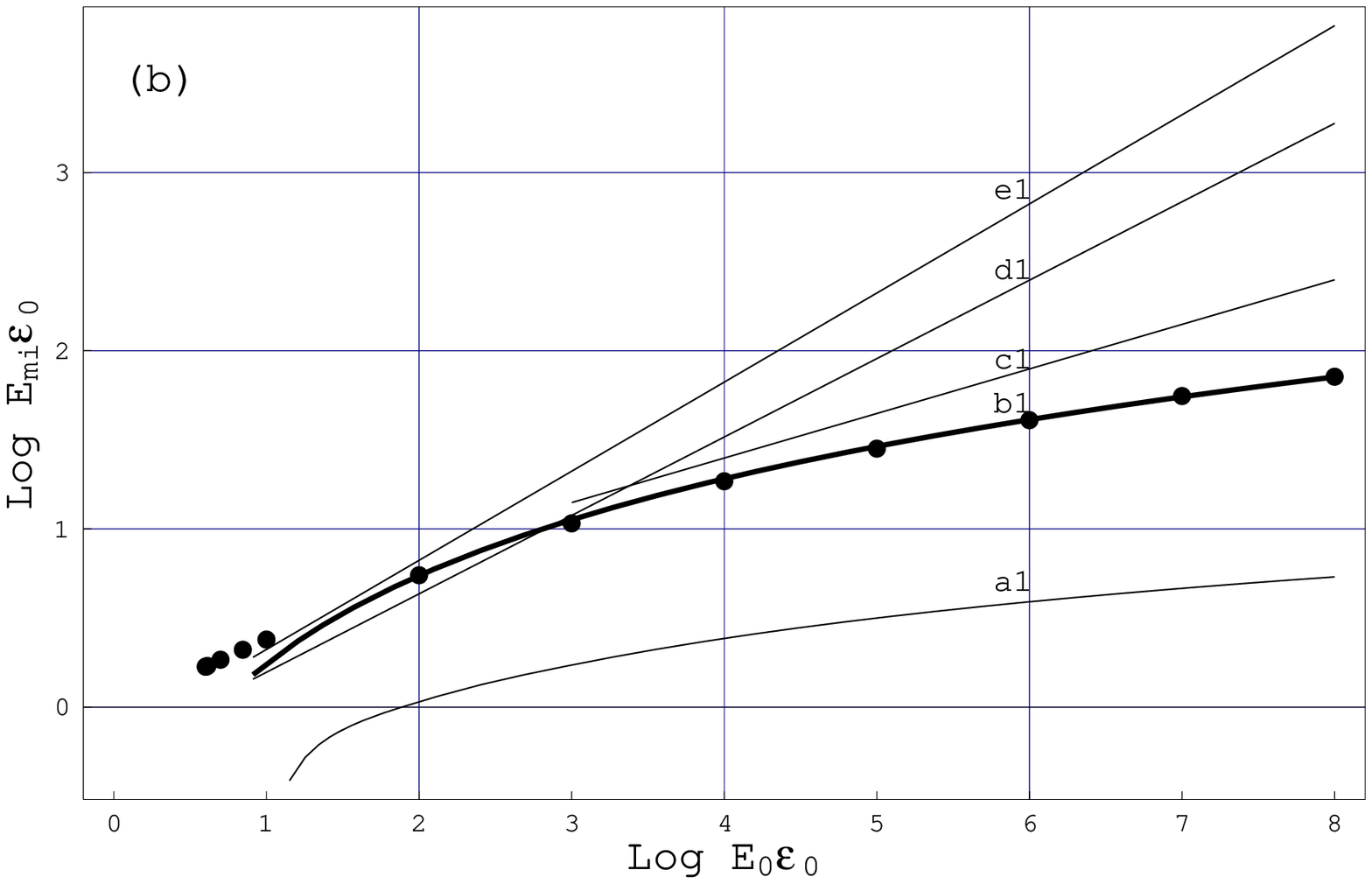} \\ 
\end{tabular}
\end{center}
\caption{(a) Plot of the quantity $f_1(s)=E_ms/E_0$ (the mean energy $E_m$
of produced particles normalized to $E_0/s$) versus ERF collision energy $s$%
. The black squares represent the results from our calculations fitted by a
cubic spline (curve($b_0$)); at $s>10^2$ curve ($b_0$) is well described by
the function $f_1(s)=0.195\ln ^2(2s)$. Curves ($e_0$) and ($a_0$) correspond
to the approximation $f_1(s)=0.71s^{1/2}$ and $f_1(s)=\ln ^2(s/4)/\sigma
_{totBH}$ based on the results of Dermer and Schlickeiser [9], and Feenberg
and Primakoff [10], respectively. (b) Plot of the quantity $\varphi
_1(s_{\perp })=E_{mi}s_{\perp }/E_0=E_{mi}\varepsilon _0$ (the mean energy
of produced particles in an isotropic and monochromatic soft photon field, $%
E_{mi}$, normalized to $E_0/s_{\perp }$) versus $s_{\perp }=\varepsilon
_0E_0 $. The black circles represent the results from our calculations. At $%
s_{\perp }\geq 10^2$ they are fitted by the function $\varphi _1(s_{\perp
})=0.195\ln ^2(2s_{\perp })$ represented by curve ($b_1$). Curves ($e_1$), ($%
d_1$), ($c_1 $) and ($a_1$) correspond to the approximations $\varphi
_1(s_{\perp })=\frac 23s_{\perp }^{1/2}$, $\varphi _1(s_{\perp
})=0.57s_{\perp }^{0.44}$, $\varphi _1(s_{\perp })=2.5s_{\perp }^{0.25}$,
and $\varphi _1(s_{\perp })=\varphi _{FP}(s_{\perp })$ based on the results
of Dermer and Schlickeiser [9], Mastichiadis \textit{et al} [2],
Mastichiadis \textit{et al} [4] and Feenberg and Primakoff [10],
respectively.}
\label{fig:EM}
\end{figure}

Thus, the knowledge of $f(s)$ allows one to determine $E_m$ for any pair of
values of $E_0$ and $s$. At fixed values of $E_0$ and $\varepsilon _0$, $f(s)
$ describes, in practice the dependence of $E_m$ on the angle of collision $%
\theta $.

The results calculated for $E_ms/E_0$ versus $s$ are represented in figure
7(a) by black squares (see also table \ref{tab2}). At $s>10^2$ they are well
described by the dependence (curve ($b_0$)):

\begin{table}[tbp]
\caption{Normalized mean energies of produced particles.}
\label{tab2}%
\begin{indented}
\lineup
\item[]\begin{tabular}{@{}*{4}{l}}
\br                              
$s,s_{\perp}$&$E_{m}s/E_{0}$&$E_{mi}s_{\perp}/E_{0}=E_{mi}\varepsilon_{0}$&$0.195\ln
^2(2s_{\perp })$\cr 
\mr
4.01&1.337&1.69&0.84\cr
4.1 &1.339&1.70&0.86\cr 
5   &1.44&1.84&1.03\cr 
7   &1.68&2.10&1.36\cr
10  &1.97&2.39&1.75\cr
$10^{2}$&5.00&5.50&5.47\cr
$10^{3}$&10.1&10.7&11.3\cr
$10^{4}$&17.7&18.5&19.1\cr
$10^{5}$&27.1&28.2&29.0\cr
$10^{6}$&39.4&40.8&41.0\cr
$10^{7}$&54.2&55.8&55.1\cr
$10^{8}$&70.0&71.5&71.2\cr 
\br
\end{tabular}
\end{indented}
\end{table}

\begin{eqnarray}
f_1(s) &=&E_m(E_0,s)s/E_0=0.195\ln ^2(2s)  \label{eq9} \\
&&  \nonumber
\end{eqnarray}
The concrete calculations are performed at $\varepsilon _0=10^{-3}$, $\theta
=\pi /2$, and various values of $E_0$ leading to various values of $%
s=\varepsilon _0E_0$. Nevertheless, the results obtained for $E_m/E_0$
should, as a whole, depend only on $s$ independently of the concrete values
of $E_0$, $\varepsilon _0$ and $\theta $. Thus, on the basis of a special
case we obtain the dependence $E_m(E_0,s)$ having a more general validity.
In figure 7(a) we have also graphically represented two other estimates of
the function $f_1(s)=E_m(E_0,s)s/E_0$ obtained by other authors. The line ($%
e_0$) corresponds to the estimate $E_m(E_0,s)=0.71E_0s^{-0.5}$obtained
analytically by Dermer and Schlickeiser \cite{derm91}. At values of $s\leq
10^2$ this line passes through our points (squares). At values of $s>10^2$
the line goes far above our points, thus predicting several orders of
magnitude higher results for $E_m$. The curve ($a_0$) corresponds to the
estimate $E_m(E_0,s)=(E_0/s)\ln ^2(s/4)/\sigma _{totBH}(s)$ obtained on the
basis of the theoretical approach developed by Feenberg and Primakoff \cite
{feen48} (see equations (19), (21), (22) and (31) in \cite{feen48}); $\sigma
_{totBH}(s)$ is an analytical approximation to the Bethe-Heitler formula for 
$\sigma _{tot}(s)$ normalized to the quantity $\alpha _fr_0^2$. It is seen
that curve ($a_0$) lies essentially below our points and predicts several
times lower results for $E_m(E_0,s)$. A reason for this is that curve ($a_0$%
) describes an approximation obtained analytically as a lower limit of the
true dependence $f_1(s)$. Another reason, that was pointed out by
Mastichiadis \textit{et al} \cite{mast86} is the neglect (in \cite{feen48})
of the recoil of the primary electron in the electron rest frame.

In order to determine the mean energy $E_{mi}$ of a particle produced by
relativistic electron-photon collision in an isotropic and monochromatic
soft photon field one should additionally average $E_m$ over the angle of
collision $\theta $. Then the expression of $E_{mi}$ is obtained in the form:

\begin{eqnarray}
E_{mi}(E_0,s_{\perp }) &=&E_0\varphi (s_{\perp }){,}  \label{eq10} \\
&&  \nonumber
\end{eqnarray}
where $\varphi (s_{\perp })=N_i(s_{\perp })/\sigma _{toti}(s_{\perp })$ is a
function of $s_{\perp }=\varepsilon _0E_0$;

\begin{eqnarray}
N_i(s_{\perp }) &=&(2\beta \varepsilon _0^2E_0^2)^{-1}\int_4^{\varepsilon
_0E_0(1+\beta )}sN(s)\rmd s{,}  \label{eq11} \\
&&  \nonumber
\end{eqnarray}

\begin{eqnarray}
\sigma _{toti}(s_{\perp }) &=&(2\beta \varepsilon
_0^2E_0^2)^{-1}\int_4^{\varepsilon _0E_0(1+\beta )}s\sigma _{tot}(s)\rmd s{.}
\label{eq12} \\
&&  \nonumber
\end{eqnarray}

The results calculated for $\varphi _1(s_{\perp })=E_{mi}(E_0,s_{\perp
})s_{\perp }/E_0=E_{mi}(E_0,s_{\perp })\varepsilon _0$ versus $s_{\perp }$
are represented in figure 7(b) by black circles (see also table \ref{tab2}).
At $s_{\perp }\geq 10^2$ they are fitted by the dependence (curve ($b_1$))

\begin{eqnarray}
\varphi _1(s_{\perp }) &=&E_{mi}(E_0,s_{\perp })\varepsilon _0=0.195\ln
^2(2s_{\perp })  \label{eq13} \\
&&  \nonumber
\end{eqnarray}
(i.e. $E_{mi}(E_0,s_{\perp })\rightarrow E_{mi}(\varepsilon _0,s_{\perp
})=(0.195/\varepsilon _0)\ln ^2(2s_{\perp })$) that has the same form as the
dependence $f_1(s)$ (equation (\ref{eq9})). There are four more curves
represented in figure 7(b), which describe some approximations of the
function $\varphi _1(s_{\perp })$ obtained by different authors. The line ($%
e_1$) corresponds to the estimate $\varphi _1(s_{\perp })=\frac 23s_{\perp
}^{1/2}$ \cite{derm91}. Certainly, it almost coincides with the line ($e_0$)
in figure 7(a), and at $s<10$ passes in immediate proximity to our points
(circles). With the increase of $s_{\perp }$ above $10^2$, the discrepancy
with our results also increases, achieving two orders of magnitude at $%
s=10^8 $. The lines ($d_1$) and ($c_1$) correspond to the approximations $%
\varphi _1(s_{\perp })=0.57s_{\perp }^{0.44}$ and $\varphi _1(s_{\perp
})=2.5s_{\perp }^{0.25}$ obtained by Mastichiadis \textit{et al} in 1986 
\cite{mast86} and 1991 \cite{mast91}, respectively. The first approximation
(line ($d_1$)) has been obtained by numerical calculations. It is near that
obtained by Dermer and Schlickeiser, and has a similar behaviour with
respect to our results. The latter approximation is obtained after
reconsidering the first one and performing improved calculations and
computer simulations. In the interval from $s=10^3$ to $s=10^8$ the line ($%
c_1$) lies just above our results. For completeness, we shall also briefly
consider the results for $\varphi _1(s_{\perp })=\varphi _{FP}(s_{\perp
})=(1/s_{\perp })\{s_{\perp }[(\ln s_{\perp }-\ln 2-1)^2+1]-[(\ln
2+1)^2+1]\}/\sigma _{totiFP}$ obtained on the basis of the approach of
Feenberg and Primakoff (curve ($a_1$)) by using the above-mentioned
equations (19), (21), (22) and (31) in \cite{feen48}; $\sigma _{totiFP}$ is
derived from equations (19) and (21) in \cite{feen48}, and is normalized to
(divided by) $\alpha _fr_0^2$. The corresponding curve (($a_1$) in figure
7(b)) almost coincides with the curve ($a_0$) in figure 7(a), thus showing
the same behaviour with respect to our results. The reasons for such a
behaviour are pointed out above. The resemblance between our results and
those of Feenberg and Primakoff is that $\varphi _1(s_{\perp })$ ($f_1(s)$)
is obtained to be proportional to $\ln ^2s_{\perp }$ and $\ln \,s_{\perp }$ (%
$\ln ^2s$ and $\ln \,s$) respectively, and not to a power function of $%
s_{\perp }$ ($s$) as in the other known approximations. Let us finally note
that the lowest estimate of $E_m\sim 2/\varepsilon _0$ was mentioned by
Blumenthal \cite{blum71}. Thus, it appears that there is a natural tendency
to improve the calculation accuracy, leading to the results obtained here.

\subsection{Primary-electron energy losses}

The primary electron energy loss rate $L_{TPP}$ (the energy lost per unit
time) due to TPP in an isotropic and monochromatic soft photon field is
given by the expression

\begin{eqnarray}
L_{TPP} &=&2cnE_0N_i(s_{\perp })=2cnE_{mi}(E_0,s_{\perp })\sigma
_{toti}(s_{\perp }){,}  \label{eq14} \\
&&  \nonumber
\end{eqnarray}
where $c$ is the speed of light and $n$ is the number of photons per unit
volume. According to the results given in table \ref{tab1} the values of $%
\sigma _{toti}(s_{\perp })$ at $s_{\perp }>10^4$ are described correctly by
the Haug formula \cite{haug81}:

\begin{eqnarray}
\sigma _{toti}(s_{\perp }) &=&\alpha _fr_0^2[\frac{28}9\ln (2s_{\perp })-%
\frac{218}{27}]{.}  \label{eq15} \\
&&  \nonumber
\end{eqnarray}
Also, as shown above, at $s_{\perp }>10^2$ the values of $%
E_{mi}(E_0,s_{\perp })$ or $E_{mi}(\varepsilon _0,s_{\perp })$ are described
correctly by equation (\ref{eq13}). Therefore, based on equations (\ref{eq13}%
)-(\ref{eq15}) we can write the following analytical expression of $L_{TPP}$
normalized to the quantity $\chi =cn\pi r_0^2/\varepsilon _0$:

\begin{eqnarray}
q_{TPP}(s_{\perp }) &=&L_{TPP}/\chi =0.386\alpha _f\ln ^2(2s_{\perp })[\ln
(2s_{\perp })-\frac{218}{84}]{.}  \label{eq16} \\
&&  \nonumber
\end{eqnarray}

The primary-electron energy loss rate $L_{ICS}$ due to inverse Compton
scattering (ICS) in an isotropic and monochromatic soft photon field
(normalized again to $\chi $) is given by \cite{blum70}:

\begin{eqnarray}
q_{ICS}(s_{\perp }) &=&L_{ICS}/\chi =\ln (4s_{\perp })-\frac{11}6{.}
\label{eq17} \\
&&  \nonumber
\end{eqnarray}

\begin{figure}[tbp]
\begin{center}
\mbox{\epsfysize=2.3in \epsffile{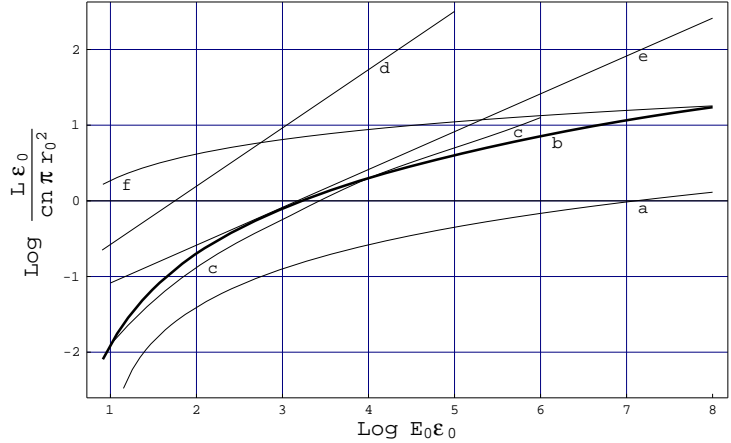}}
\end{center}
\caption{Plot of the quantity $q_{TPP}=L_{TPP}/\chi $ (TPP energy loss rate $%
L_{TPP}$ normalized to $\chi =cn\pi r_0^2$) versus $s_{\perp }=\varepsilon
_0E_0$, compared with the analogous quantity $q_{ICS}=L_{ICS}/\chi $
concerning ICS energy loss rate $L_{ICS}$ (curve ($f$)). Curve ($b$) is
obtained on the basis of precise calculations performed in this work. Curves
($c$), ($d$), ($e$) and ($a$) represent some approximations based on the
results of Mastichiadis \textit{et al} [4], Mastichiadis \textit{et al} [2],
Dermer and Schlickeiser [9] and Feenberg and Primakoff [10], respectively.}
\label{fig:EL}
\end{figure}

The normalized losses $q_{TPP}$ and $q_{ICS}$ versus $s_{\perp }$ are
compared graphically in figure 8. There, curve ($f$) represents the
dependence $q_{ICS}(s_{\perp })$ given by equation (\ref{eq17}). Curve ($b$)
is obtained on the basis of precise numerical calculations performed in this
work. It can be seen that, according to our results, TPP losses become
prevalent and increase above ICS losses at values of $s_{\perp }$ exceeding
a threshold $s_{th}\sim 10^8$. Certainly, this threshold is considerably
higher (five orders of magnitude) than another threshold $s^{*}=250$ at
which the interaction lengths of TPP and ICS become equal \cite{mast91}. An
estimate of $s_{th}$ derived from the equality $q_{TPP}(s_{\perp
})=q_{ICS}(s_{\perp })$ (by using equations (\ref{eq16}) and(\ref{eq17})) is 
$s_{th}\simeq 1.6\times 10^8$.

The estimate of $q_{TPP}(s_{\perp })=(5/\pi )\alpha _fs_{\perp }^{1/4}[\frac{%
28}9\ln (2s_{\perp })-\frac{218}{27}]$ obtained by Mastichiadis \textit{et al%
} in 1991, is shown by curve ($c$). The threshold predicted in this case is $%
s_{th}\simeq 10^6$. Curve ($e$) represents the estimate $q_{TPP}(s_{\perp })=%
\frac{32}9\alpha _fs_{\perp }^{1/2}$obtained by Dermer and Schlickeiser \cite
{derm91}. It predicts a threshold $s_{th}\simeq 2\times 10^5$. Two more
estimates of the dependence $q_{TPP}(s_{\perp })$ obtained from the results
of \cite{feen48} and \cite{mast86} are illustrated by curves ($a$) and ($d$%
), and give unrealistically high and low thresholds, respectively. So, the
consideration performed here, of the results obtained by different authors
for the primary-electron energy loss rate due to TPP, confirms the existence
of a tendency to a permanent improvement of the calculation accuracy.
Because of the efforts to overcome the calculation difficulties and
additionally increase the calculation precision, one can accept the results
obtained here as reliable and accurate. They show that the electron energy
losses due to TPP, e.g. in cascading processes occurring in pulsars (\cite
{mast87}, \cite{derm91}, \cite{stur95}) or in photon background field (\cite
{mast86}, \cite{mast91}, \cite{lee98}), are lower than those formerly
predicted. The differences between the results for DDCS, SDCS, $E_m(E_0,s),$%
and $E_{mi}(E_0,s_{\perp })$ obtained here and those obtained in other works
might lead to differences in the results from modelling electron-photon
cascading in a soft photon field. Certainly, the detailed simulations now in
progress will reveal the influence on the final results of the differences
and factors discussed in this paper.

\section{Conclusion}

In order to develop a Monte Carlo code for three-dimensional modelling of
TPP, we have undertaken a series of systematic precise calculations of DDCS
and SDCS with respect to the produced electron and positron energies, as
well as of the total cross-sections of TPP in the laboratory frame. The
behaviour of the mean produced-particle energies has also been investigated
in detail. To avoid crucial irregularities and sharp variations of the
integrand, and extremely short integration intervals in the expressions of
DDCS, SDCS, total cross-section, and mean produced-particle energy, we have
used some appropriate mathematical approaches such as suitable changes of
variables, optimum-power spline technique etc. These approaches lead to
simpler and regular expressions of the cross-sections as well as to stable,
accurate and accelerated calculation procedures. In addition, the use of the
modern powerful program code \textit{Mathematica}, working with arbitrary
precision numbers, allowed us to obtain reliable high-precision results.
Thus, the DDCS, SDCS and total cross-section have been computed for a
variety of initial and final parameters characterizing TPP. The results for
the total cross-section are in excellent agreement with ones obtained by
other authors. However, there are some discrepancies in the results for DDCS
and SDCS that might lead to differences in the results from modelling.

The mean produced-particle energy $E_m$ is analytically confirmed (in a
general form) to be proportional to the incoming electron energy $E_0$ and
to a function $f(s)$ of the collisional parameter $s$ only, i.e. $%
E_m=E_m(E_0,s)=E_0f(s)$. It is also confirmed analytically that in an
isotropic and monochromatic soft photon field the mean produced-particle
energy $E_{mi}$ (averaged over the angle of collision $\theta $) is
proportional to $E_0$ and a function $\varphi (s_{\perp })$ of the product $%
s_{\perp }=\varepsilon _0E_0$, i.e. $E_{mi}=E_{mi}(E_0,s_{\perp
})=E_0\varphi (s_{\perp })$. Such a general behaviour established of $\ E_m$
and $E_{mi}$ is in agreement with the results of other authors (\cite{mast86}%
, \cite{mast91}, \cite{derm91}, \cite{feen48}). However, there are some
essential differences that would also lead to different results from
modelling. So, the mean produced-particle energy $E_m$ or $E_{mi}$ obtained
here is proportional (apart from $E_0/s$ or $E_0/s_{\perp }$) to $\ln ^2(2s)$
or $\ln ^2(2s_{\perp })$, respectively (see equations (\ref{eq9}) and (\ref
{eq13})). At the same time, some earlier investigations of this question
have led to a proportionality to a power function such as $s^{0.5}$ \cite
{derm91}, $s_{\perp }^{0.44}$ \cite{mast86} or $s_{\perp }^{0.25}$ \cite
{mast91}. The indicated differences in the determination of $E_{mi}$ lead to
differences in the determination of that threshold level of $s_{\perp
}=s_{th}$ above which the primary-electron energy losses due to TPP become
prevalent over the ICS energy losses. It is shown here that the value of $%
s_{th}\sim 10^8$, that differs from the values of $s_{th}\sim 10^6$ \cite
{mast91} or $2\times 10^5$ \cite{derm91} obtained formerly. The
last-mentioned result means that there has been some overestimation of the
role of TPP energy losses in some astrophysical studies (\cite{mast86}-\cite
{mast91}, \cite{derm91}, \cite{stur95}, \cite{lee98}).

\section*{Appendix. Expression of the integrand function $X$.}

The integrand function $X=X(E_0,\varepsilon _0,s,E_{+},E_{-},\xi ,\eta ,\phi
_{+})$ is expressible in the following (possibly the only one) viewable and
compact form that facilitates the programming and the calculations to be
performed (see also \cite{mork71}, \cite{mork73}, \cite{motz69}): 
\[
X=X_U+X_V+X_W, 
\]
where

\[
\fl X_U=U+S_1U+S_2U+S_3U+S_2S_1U+S_3S_1U+S_3S_2U+S_3S_2S_1U, 
\]

\[
\fl X_V=V+S_1V+S_2V+S_3V+S_2S_1V+S_3S_1V+S_3S_2V+S_3S_2S_1V, 
\]

\[
\fl X_W=W+S_1W+S_2W+S_3W+S_2S_1W+S_3S_1W+S_3S_2W+S_3S_2S_1W; 
\]

$S_1$, $S_2$, and $S_3$ denote the substitution: 
\begin{eqnarray*}
S_1 &=&\overline{k}\leftrightarrow -\overline{k},\quad \ \overline{p_0}%
\leftrightarrow \overline{p_r},\quad \ \overline{p_{-}}\leftrightarrow -%
\overline{p_{+}}, \\
S_2 &=&\overline{p_0}\leftrightarrow -\overline{p_{+}}, \\
S_3 &=&\overline{p_r}\leftrightarrow \overline{p_{-}},
\end{eqnarray*}
where (with respect to the laboratory frame) $\overline{k}=\{\overrightarrow{%
k},\varepsilon _0\}$ is the four-vector of the incoming photon; $\overline{%
p_0}=\{\overrightarrow{p_0},E_0\}$ is the four-vector of the incoming
primary electron, and $\overline{p_r}=\{\overrightarrow{p_r},E_r\}$ is the
four-vector of the recoiling primary electron; $\overline{p_{+}}=\{%
\overrightarrow{p_{+}},E_{+}\}$ and $\overline{p_{\_}}=\{\overrightarrow{%
p_{-}},E_{-}\}$ are the four-vectors of the produced positron and electron,
respectively; $\overrightarrow{k_0}$, $\overrightarrow{p_0}$, $%
\overrightarrow{p_r}$, $\overrightarrow{p_{+}}$ and $\overrightarrow{p_{-}}$
are the corresponding three-component momentum vectors, and $\varepsilon _0$%
, $E_0$, $E_r$, $E_{+}$, and $E_{-}$ are the corresponding energy values.
The module of each three-component vector $\overrightarrow{v}$ is denoted by 
$v$.

The expressions of $U$, $V$, and $W$ are: 
\begin{eqnarray*}
\fl U =\case12(1/(1+\tau _1)^2)\{(1/k_3^2)[-k_3(k_1\tau _2+k_0\sigma
_3)+\tau _1k_3-\tau _2\sigma _1-\tau _3\sigma _3 \\
\lo+k_1\tau _2+k_0\sigma _3-k_2k_3+k_2+\tau _1+2k_3-\sigma _2+2] \\
\lo+(1/(k_2k_3))[\sigma _2(k_1(\tau _2+\tau _3)-\sigma _1\tau _2-\sigma
_3\tau _3)+k_2(\sigma _1\tau _2+\sigma _3\tau _3-2\tau _3\sigma _1) \\
\lo+\sigma _2(\tau _1-\sigma _2+2k_2)-k_0k_1-\tau _1k_2+2\sigma _2-k_2]\},
\end{eqnarray*}
\begin{eqnarray*}
\fl V =\case14(1/((1+\tau _1)(1-\sigma _2)))\{(1/(k_0k_3))[2(k_0-k_3-2\tau
_3)+k_0(k_1+\tau _1+\sigma _1-\sigma _2+\sigma _3) \\
\lo+k_3(-k_2-\tau _1+\tau _2+\sigma _2-\sigma _3)+\tau _3(-k_1-k_2+2\sigma
_2-2\tau _1) \\
\lo+k_0(\sigma _1(-\sigma _2-\tau _2)-2\sigma _3(k_3+\tau _3))+k_3(\tau
_1\tau _2+\sigma _1\tau _2+2\sigma _3\tau _3) \\
\lo+\tau _3(2(\tau _2\sigma _1+\sigma _3\tau _3)-k_1\tau _2+k_2\sigma
_1)]\} \\
\lo-\case14(1/((1+\tau _1)(1-\sigma _2)))\{(1/(k_0k_2))[2(k_0-k_2-2\tau _2)
\\
\lo+k_0(k_1+\tau _1+\sigma _3-\sigma _2+\sigma _1)+k_2(-k_3-\tau _1+\tau
_3+\sigma _2-\sigma _1) \\
\lo+\tau _2(-k_1-k_3+2\sigma _2-2\tau _1)+k_0(\sigma _3(-\sigma _2-\tau
_3)-2\sigma _1(k_2+\tau _2)) \\
\lo+k_2(\tau _1\tau _3+\sigma _3\tau _3+2\sigma _1\tau _2)+\tau _2(2(\tau
_3\sigma _3+\sigma _1\tau _2)-k_1\tau _3+k_3\sigma _3)]\},
\end{eqnarray*}
\begin{eqnarray*}
\fl W =\case18(1/((1+\tau _1)(\sigma _1-1)))\{(2/k_2^2)[2k_1k_2\tau
_3+k_2(-k_0-k_1+k_3-\tau _1+\tau _3+\sigma _1) \\
\lo+2\tau _3(\sigma _3-k_1)+k_0+k_1-2k_2-k_3-\tau _1-\tau _2+\tau
_3+\sigma _1+\sigma _2-\sigma _3-2] \\
\lo+(1/(k_2k_0))[2(\sigma _3(k_2\tau _3+k_3\tau _2-k_0(\sigma _2+\tau
_3)+2\tau _2\tau _3)+\tau _3(k_2\tau _1-k_1\tau _2)) \\
\lo+2k_0(k_1-k_3+\tau _1+\tau _2-\tau _3-\case12\sigma _1-\sigma _2+\sigma
_3)+k_1(2k_3-\tau _3+\sigma _2) \\
\lo+k_2(2\tau _3+\sigma _1-2\tau _1-2\tau _2)+k_3(\tau _1-\sigma _3)-2\tau
_2(\tau _1+\tau _2-\tau _3-\sigma _1-\sigma _2+\sigma _3) \\
\lo+2(k_1-k_3)+k_0-k_2-4\tau _2] \\
\lo+(1/(k_2k_3))[2(\sigma _3(k_3\tau _2-k_3\tau _3-k_2\tau _3-k_0\sigma
_2+2\sigma _2\tau _3)+\tau _3(\sigma _1k_2-k_1\sigma _2)) \\
\lo+k_0(2k_1+2k_3-\sigma _1-\sigma _3)+k_1(2k_3+\tau _2+\tau _3)+k_2(\tau
_1-2\tau _3-2\sigma _1-2\sigma _2) \\
\lo+k_3(\tau _1+2(\tau _2-\tau _3-\sigma _1-\sigma _2+\sigma _3)) \\
\lo+2\sigma _2(-\tau _1-\tau _2+\tau _3+\sigma _1+\sigma _2-\sigma
_3)+(-2k_0-2k_1+k_2+k_3-4\sigma _2)] \\
\lo+(1/(k_0k_3))[4\sigma _3(k_3+\tau _3)(k_0-\tau _3)+2\tau _3(\tau
_1+\tau _2-\tau _3-\sigma _1-\sigma _2+\sigma _3) \\
\lo+k_0(-2\tau _2+2\tau _3-2k_2-3\sigma _3)+k_1(2k_2-\tau _2+2\tau
_3+\sigma _2) \\
\lo+k_2(2k_3+\tau _1+2\tau _3-\sigma _1)+k_3(-2\tau _3-2\sigma _2+3\sigma
_3) \\
\lo+2k_1-2k_2+3k_3-3k_0+4\tau _3]\}.
\end{eqnarray*}

They are functions of the invariant products: 
\begin{eqnarray*}
k_0 &=&\overline{p_0}.\overline{k}=-s,\qquad \tau _1=\overline{p_0}.%
\overline{p_r}\,,\qquad \sigma _1=\overline{p_{+}}.\overline{p_r}\,, \\
k_1 &=&\overline{p_r}\,.\overline{k},\qquad \qquad \;\tau _2=\overline{p_0}.%
\overline{p_{-}},\qquad \sigma _2=\overline{p_{+}}.\overline{p_{-}}, \\
k_2 &=&\overline{p_{-}}.\overline{k},\qquad \qquad \;\tau _3=\overline{p_0}.%
\overline{p_{+}},\qquad \sigma _3=\overline{p_r}.\overline{p_{-}}, \\
k_3 &=&\overline{p_{+}}.\overline{k},
\end{eqnarray*}
which are connected, because of the energy-momentum conservation laws, by
the relations:

\begin{eqnarray*}
\sigma _3 &=&k_0+1-\sigma _1-\sigma _2, \\
\sigma _2 &=&k_0-k_1-\tau _1, \\
\sigma _1 &=&k_0-k_2-\tau _2, \\
\tau _1 &=&k_0-1-\tau _3-\tau _2, \\
k_1 &=&k_0-k_3-k_2;
\end{eqnarray*}
\ where

\[
\fl k_0=-E_0\varepsilon _0(1-\beta \cos \theta )=-s\qquad \mathrm{(section\
2.1)}, 
\]

\[
\fl k_2=p_{-}k[y\cos \theta _k+(1-y^2)^{1/2}\cos \phi _{-}\sin \theta
_k]-E_{-}\varepsilon _0, 
\]

\[
\fl k_3=p_{+}k[x\cos \theta _k+(1-x^2)^{1/2}\cos \phi _{+}\sin \theta
_k]-E_{+}\varepsilon _0, 
\]

\[
\fl \tau _2=p_{-}p_0[y\cos \theta _0-(1-y^2)^{1/2}\cos \phi _{-}\sin \theta
_0]-E_{-}E_0, 
\]

\[
\fl \tau _3=p_{+}p_0[x\cos \theta _0-(1-x^2)^{1/2}\cos \phi _{+}\sin \theta
_0]-E_{+}E_0, 
\]

\[
\fl \beta =p_0/E_0, 
\]

\[
\fl \cos \theta _0=\cos \theta \cos \theta _k+\sin \theta \sin \theta _k, 
\]

\[
\fl \sin \theta _0=\sin \theta \cos \theta _k-\cos \theta \sin \theta _k, 
\]

\[
\fl \sin \theta _k=(1-\cos ^2\theta _k)^{1/2}, 
\]

\[
\fl \cos \theta _k=(k+p_0\cos \theta )/P_{tot}, 
\]

\[
\fl \phi _{-}=\phi _{+}-\alpha , 
\]
\begin{eqnarray*}
\fl \alpha=[(1+s-E_{tot}(E_{+}+E_{-})+E_{+}E_{-}+P_{tot}(p_{+}x+p_{-}y) \\
\lo-p_{+}p_{-}xy)/(p_{+}p_{-}(1-x^2)^{1/2}(1-y^2)^{1/2})],
\end{eqnarray*}
\[
\fl P_{tot}=(p_0^2+k^2+2p_0k\cos \theta )^{1/2}, 
\]

\[
\fl p_k=\varepsilon _0,\quad p_0=(E_0^2-1)^{1/2}, 
\]

\[
\fl p_{-}=(E_{-}^2-1)^{1/2},\quad p_{+}=(E_{+}^2-1)^{1/2}. 
\]

With $\theta _0$ and $\theta _k$ we have, respectively, denoted the polar
angles of the incoming electron and photon momenta with respect to the axis
along $\overrightarrow{P_{tot}}=\overrightarrow{p_0}+\overrightarrow{k}$. In
this case $\theta _0+\theta _k=\theta $, where $\theta $ is the angle
between $\overrightarrow{p_0}$ and $\overrightarrow{k}$, i.e. the angle of
collision. The azimuthal angles of the produced electron and positron are $%
\phi _{-}$ and $\phi _{+}$, respectively . They are accounted for in such a
way that $\phi _k=0$ and $\phi _0=\pi $, where $\phi _k$ and $\phi _0$ are
the azimuthal angles of the incoming photon and electron, respectively \cite
{mast86}.

The variables $\xi $ and $\eta $ are introduced by the substitutions
(section 2.1, equations (\ref{eq2a}) and (\ref{eq2b})):

\begin{eqnarray*}
x &=&x(\xi )=\xi /l+x_{\min }\quad (l>1), \\
y &=&y(\xi ,\eta )=[y_{\max }(x)+y_{\min }(x)\eta ^2]/(1+\eta ^2).
\end{eqnarray*}

The remaining designations are given in section 2.1.

\section*{Acknowledgements}

Many thanks are due to Dr H Vankov for his helpful assistance and comments
on this study. The authors acknowledge the financial assistance of the
Bulgarian National Scientific Fund under contract F-460.

\section*{References\ }

\end{document}